\documentclass[%
 reprint,
 amsmath,
 amssymb,
 aps,
 pra,
 showpacs
]{revtex4-1}

\usepackage{graphicx}
\usepackage{dcolumn}
\usepackage{bm}
\usepackage{bbold}
\usepackage[export]{adjustbox}
\usepackage{amsmath,amsfonts,mathtools}
\usepackage{dsfont}
\usepackage{amsmath}
\usepackage{amssymb}
\usepackage{makeidx}
\usepackage{graphicx}
\usepackage{color}

\newcommand{\hide}[1]{}

\newcommand{\lcase}{\left\{\begin{array}{ll}}
\newcommand{\rcase}{\end{array}\right.}
\newcommand{\ear}{\end{array}}
\newcommand{\bal}{\begin{align}}
\newcommand{\eal}{\end{align}}
\newcommand{\bma}{\begin{pmatrix}}
\newcommand{\ema}{\end{pmatrix}}
\newcommand{\beq}{\begin{equation}}
\newcommand{\eeq}{\end{equation}}
\newcommand{\bel}[1]{\begin{equation}\label{eq:#1}}
\newcommand{\eel}{\end{equation}}
\newcommand{\bea}{\begin{eqnarray}}
\newcommand{\eea}{\end{eqnarray}}
\newcommand{\beaNN}{\begin{eqnarray*}}
\newcommand{\eeaNN}{\end{eqnarray*}}

\newcounter{lecture}

\renewcommand{\hide}[1]{}

\begin{document}

\preprint{APS/123-QED}

\title{Resonant Coupling Between Electromagnetic Waves and Protein Conformational Dynamics Revealed by Molecular Dynamics Simulations}
\author{Jiafei Chen}
\thanks{These authors contributed equally to this work.}
\affiliation{%
The College of Medical Technology, Shanghai University of Medicine \& Health Sciences, Shanghai 200100, China
}
\author{Yuanyuan Feng}
\thanks{These authors contributed equally to this work.}
\affiliation{%
The College of Medical Technology, Shanghai University of Medicine \& Health Sciences, Shanghai 200100, China
}
\author{Jingzhi Feng}
\thanks{These authors contributed equally to this work.}
\affiliation{%
The College of Medical Technology, Shanghai University of Medicine \& Health Sciences, Shanghai 200100, China
}
\author{Xinyun Zhang}
\thanks{These authors contributed equally to this work.}
\affiliation{%
The College of Medical Technology, Shanghai University of Medicine \& Health Sciences, Shanghai 200100, China
}
\author{Jinzhen Zhu}
\email{zhujinzhenlmu@gmail.com}
\affiliation{%
Shanghai Artificial Intelligence Laboratory, 129 Longwen Road, Shanghai, China
}%
\author{Qingmeng Xu}
\email{a956424513@163.com}
\affiliation{%
The College of Medical Technology, Shanghai University of Medicine \& Health Sciences, Shanghai 200100, China
}
\begin{abstract}
The biological effects of electromagnetic fields on proteins remain controversial beyond well-established thermal mechanisms, particularly with respect to frequency-dependent responses. Here, we propose that electromagnetic waves can modulate protein conformation through resonant coupling with intrinsic protein dynamics. Molecular dynamics simulations were employed to characterize spontaneous conformational fluctuations in the absence of external fields, and a tiered screening strategy combined with fast Fourier transform analysis was used to identify dominant intrinsic frequencies associated with periodically fluctuating non-covalent atom or residue pairs. Oscillating external electric fields were subsequently applied at resonant and off-resonant frequencies to evaluate conformational responses across diverse protein systems. The results demonstrate that resonant excitation induces significantly enhanced backbone conformational deviations compared to off-resonant conditions, with the effect becoming more pronounced in structurally flexible and multichain proteins. These findings provide atomistic evidence for frequency-specific resonance between electromagnetic fields and protein conformational dynamics, offering mechanistic insight into frequency-dependent electromagnetic effects and a computational framework for electromagnetic wave–based modulation of protein function.
\end{abstract}

\maketitle

\section{\label{sec:intro}Introduction}
With the rapid proliferation of wireless technologies, human exposure to radiofrequency electromagnetic fields (RF-EMF) has become nearly ubiquitous. Sources ranging from mobile networks (4G/5G) and satellite transmissions to high-voltage power lines have fundamentally altered the electromagnetic environment~\cite{ICNIRP2020}. Global statistics indicate that mobile telephone subscriptions exceeded 9.1 billion by 2024, a figure approaching the total human population~\cite{WikiMobilePhone}. This unprecedented level of widespread exposure has intensified scientific and public scrutiny regarding the potential biological and health effects of RF-EMF, a subject of rigorous investigation since the late twentieth century~\cite{Repacholi1998,Roosli2010}.

Proteins are the primary functional workhorses of the cell, executing essential biological roles across all living organisms~\cite{BrandenTooze1999,PetskoRinge2004}. Composed of L-$\alpha$-amino acids polymerized via peptide bonds~\cite{Kyte1991}, proteins are organized into a sophisticated structural hierarchy encompassing primary, secondary, tertiary, and quaternary levels~\cite{BrandenTooze1999,MaoHua2012}. The primary structure—the linear amino acid sequence—acts as a fundamental blueprint that dictates folding into secondary elements, such as $\alpha$-helices and $\beta$-sheets, primarily through non-covalent interactions~\cite{Kyte1991,PetskoRinge2004}. These elements further collapse into a compact, globular tertiary structure defined by the spatial arrangement of secondary motifs and their connecting loops~\cite{BrandenTooze1999}. Finally, multiple polypeptide subunits may assemble into quaternary complexes, a process vital for the functional regulation of many enzymatic and signaling systems~\cite{PetskoRinge2004,Kyte1991}.

The biological utility of a protein is inextricably linked to its specific three-dimensional conformation; thus, the folding process is both essential and stringently regulated~\cite{Anfinsen1973,Dobson2003}. Proteins achieve their native state through a delicate thermodynamic equilibrium of non-covalent forces, including the hydrophobic effect, van der Waals interactions, hydrogen bonding, and electrostatic forces~\cite{Dill1990,BrandenTooze1999}. However, folding is entropically unfavorable as it severely restricts the conformational freedom of the polypeptide backbone (dihedral angles $\phi$ and $\psi$) and side chains ($\chi$ angles)~\cite{DillMacCallum2012,PetskoRinge2004}. This sensitivity makes the folding process highly susceptible to external perturbations, including temperature, pH, chemical denaturants, molecular crowding, and exogenous electric or magnetic fields~\cite{Ellis2001,Wang2016}. Even minor structural disruptions—stemming from point mutations or environmental stress—can lead to misfolding or denaturation, underlying a spectrum of proteopathic diseases and cellular dysfunction~\cite{Dobson2003,ChitiDobson2006}.

Crucially, protein function is defined not only by a static structure but also by complex conformational dynamics under physiological conditions~\cite{HenzlerWildman2007,Frauenfelder1991}. While techniques such as X-ray crystallography and cryo-electron microscopy provide high-resolution "snapshots," they often fail to capture the continuous fluctuations and dynamic responses to subtle environmental stimuli~\cite{Karplus2002,Dror2012}. Molecular dynamics (MD) simulations bridge this gap by numerically solving Newton’s equations of motion, tracking the trajectories of all atoms in a protein-solvent system from femtoseconds to microseconds~\cite{McCammon1987,Karplus2005}. This atomistic approach reveals intrinsic vibrational modes and flexibility distributions, providing a powerful framework to study how exogenous physical stimuli, such as electromagnetic fields, influence protein conformation~\cite{Hunenberger1999,Wang2014}.

The potential for electromagnetic energy to modify protein structure has long been a subject of interest. Epidemiological concerns date back to 1979, when a correlation was reported between residential proximity to high-voltage lines and increased childhood cancer risk~\cite{WertheimerLeeper1979}. Conversely, external electric fields (EEFs) are increasingly utilized in therapeutic contexts. For instance, the efficacy of electromagnetic waves in disrupting amyloid plaques is known to be frequency-dependent; simulations suggest that oscillating EEFs (Os-EEFs) in the GHz range can decompose plaques, whereas THz frequencies are less effective. However, the biophysical basis for this frequency dependence remains largely unexplained. While some models propose that pulsed microwave radiation affects proteins primarily through localized thermal effects~\cite{BohrBohr2000}, others suggest direct, non-thermal interactions with the macromolecular structure~\cite{Porcellietal1997,Taylor1981,Byusetal1984}. Despite these hypotheses, experimental and theoretical support for non-thermal RF-EMF effects remains insufficient.

In this study, we propose that electromagnetic waves induce conformational changes through resonance with specific non-covalently linked atom or residue pairs within the protein. Using MD simulations, we tracked protein conformational evolution and implemented a Python-based tiered screening algorithm to identify periodically fluctuating pairs. Coupled with Fast Fourier Transform (FFT) analysis~\cite{Press2007,CooleyTukey1965}, this approach allowed for the systematic exploration of candidate resonance frequencies in both single-chain and multichain systems. Our results provide atomistic evidence for frequency-specific resonance phenomena, offering new insights into the mechanical coupling between electromagnetic radiation and protein dynamics.

\section{Computational Methods}
\begin{figure*}[htbp]
\centering
\includegraphics[width=0.8\textwidth]{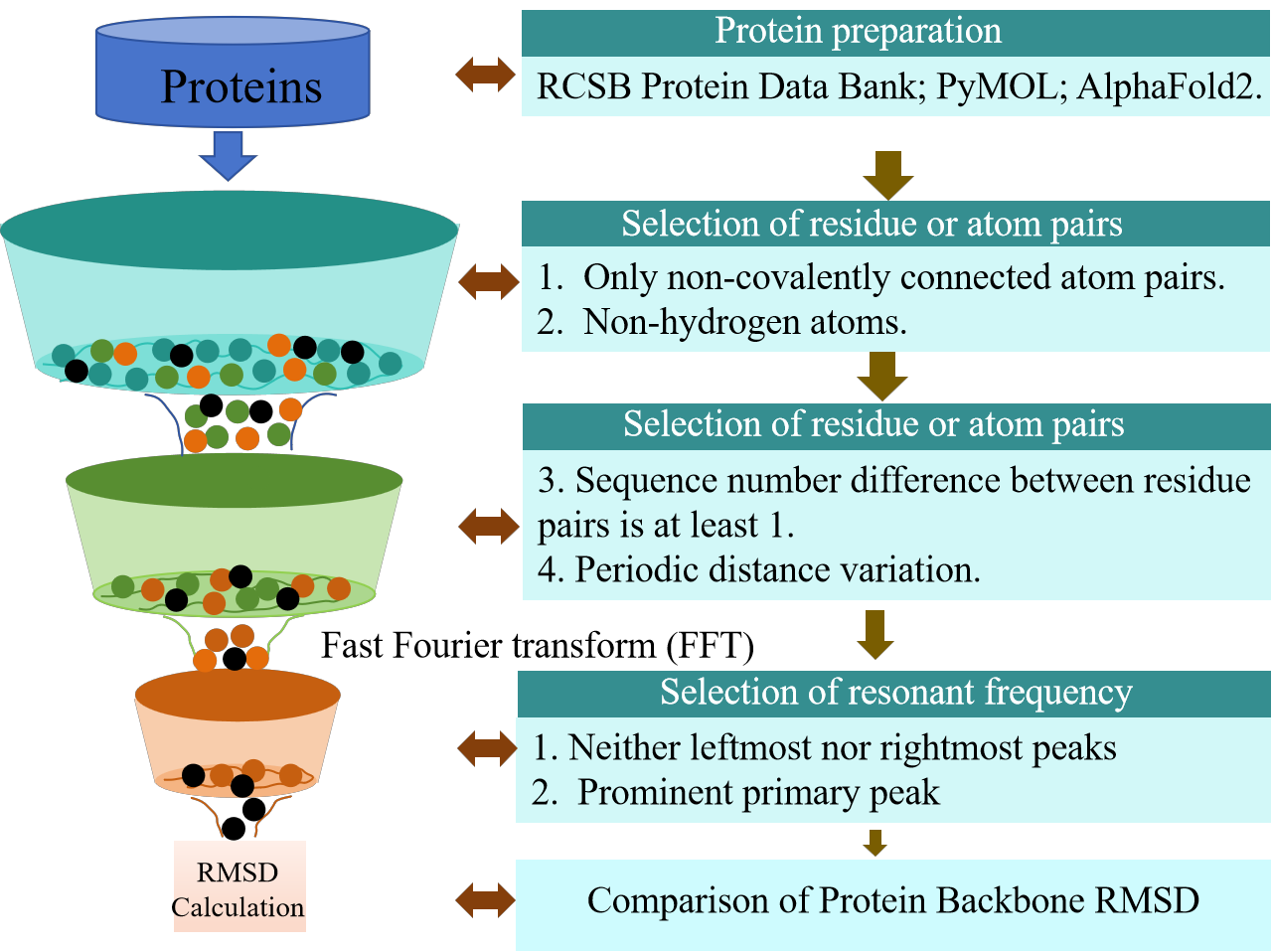}
\caption{The workflow for predicting protein conformational changes under resonant frequency.
}
\label{fig:workflow}
\end{figure*}

The procedure can be generally divided into four steps: protein preparation, selection of residue or atom pairs, selection of resonant frequency and RMSD analysis, as depicted in Figure~\ref{fig:workflow}.
\subsection{Structural Modeling and System Setup}
The target protein structures were obtained from the RCSB Protein Data Bank (https://www.rcsb.org/) database. This starting protein structure underwent preprocessing using molecular visualization software (e.g., PyMOL~\cite{PyMOL} or VMD~\cite{VMD1996}) to remove crystallographic water molecules, extraneous ions, and any pre-existing ligand molecules, thereby retaining only the complete protein chain and yielding a purified target protein. For proteins with incomplete or unavailable experimental structures, AlphaFold2~\cite{AlphaFold2,AlphaFoldDatabase2024} was employed for structure prediction, and only models with an average pLDDT ≥ 90 were selected as high‑confidence structures. Each processed protein was then placed in a cubic periodic boundary box, where the box edge length was set to exceed the maximum dimension of the protein by 3.0 nm to ensure adequate solvation space.
\subsection{Molecular Dynamics Simulations}
The first step in preparing the system is to define the topology of the protein, set protonation states of side chains and termini, specify disulfide linkages, and build in any missing hydrogen atoms. These processes are carried out in force fields, Amber99sb~\cite{Amber99sb,Amber_Original}, provided with GROMACS~\cite{GROMACS2015,GROMACS_Original}. Water molecules were modeled with the TIP3P~\cite{Jorgensenetal1983} water model. To reach a suitably low potential energy, equilibration was performed sequentially under the NVT (constant number of particles, volume, and temperature) and NPT (constant number of particles, pressure, and temperature) ensembles.
Molecular dynamics simulations were carried out for a total duration of 10-20 nanoseconds (ns) with an integration time step of 2 femtoseconds.  
\subsection{Time-Series Analysis and Pair Screening}
The generated atomic coordinate trajectory files were subjected to desolvation, periodic boundary condition correction, and centering. To screen for atom or residue pairs with potential resonance characteristics, the following criteria were applied:
\begin{itemize}
    \item The atom pair does not consist of atoms directly connected by chemical bonds within the same residue;
    \item Both atoms in the pair are non-hydrogen atoms;  
    \item For residue pairs, the sequence number difference between the two residues is at least 1;  
    \item The distance variation of the atom pair or residue pair exhibits periodic characteristics. 
\end{itemize}

\subsection{Spectral Processing and Frequency Extraction}
Using the fast Fourier transform algorithm (FFT), the preprocessed time-series data of relative distances between atom or residue pairs was converted from the time domain to the frequency domain. The frequencies corresponding to distinct peaks in the resulting spectra were selected for further screening of potential resonance frequencies.
\subsection{Conformational Response Analysis}
Based on the screened frequencies, an external electric field was applied to the protein at three distinct frequency conditions: one-tenth of the selected frequency, ten times the selected frequency, and the original frequency itself. The root-mean-square deviation (RMSD) of the protein backbone was computed and compared across these conditions to evaluate the influence of resonance on protein conformational dynamics.
We fully admit that the field intensity employed by us is significantly higher than the clinically used external electric fields (EEFs). Nevertheless, we utilize higher intensity because it allows us to perform MD simulations for reasonably long time scales and provides meaningful mechanistic insights. English et al. demonstrate that the observation of tangible field effects at higher field strengths requires rather short simulation time scales.47,48 It has been supported by experimental evidence that peptides exhibit similar behavior when the field strength is reduced to clinically relevant levels. 
\subsection{Software and Tools}
Structural preprocessing: PyMOL, Structure prediction: AlphaFold2, Molecular dynamics simulations: GROMACS, Trajectory analysis: MDAnalysis, Signal processing and plotting: Python.

\section{Result and discussion} 

\subsection{The selected proteins}
\begin{figure*}[htbp]
\centering
\includegraphics[width=0.8\textwidth]{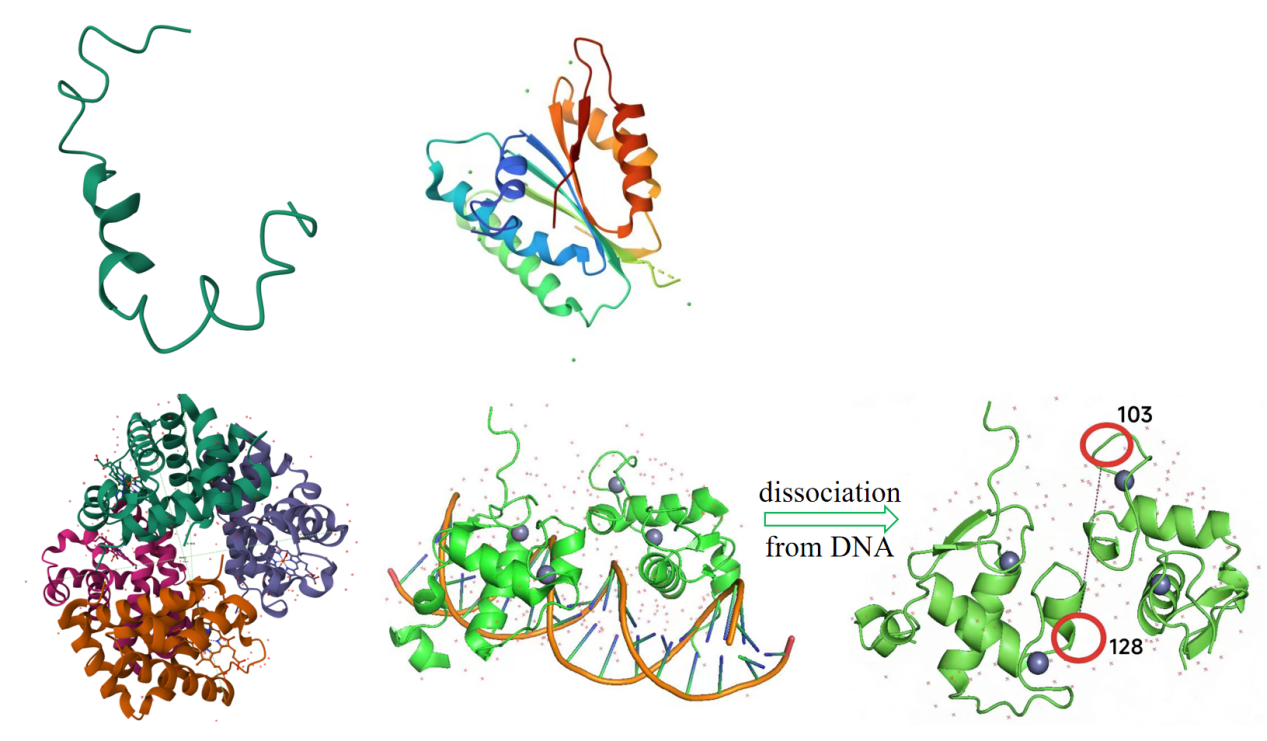}
\caption{The selected proteins. (a) short single-chain protein (PDB: 6SZF), (b) longer single-chain protein (PDB: 3A1F), (c) multichain protein (PDB: 1GZX), (d) Structurally flexible multichain protein (PDB: 2HAN).
}
\label{fig:protein-structure}
\end{figure*}

To validate the effectiveness of the method, four distinct proteins were selected: single-chain A$\beta$42 peptide (PDB ID: 6SZF), longer single‑chain NOX2 (modeled via AlphaFold2, based on PDB template 3A1F), heterotetrameric human hemoglobin (PDB ID: 1GZX) and DNA‑binding domain of the ecdysone receptor (PDB ID: 2HAN), which was studied in its DNA-free, conformationally flexible state.
Alzheimer’s disease is the most frequent neurodegenerative disease and the leading cause of dementia worldwide. Amyloid-$\beta$ 42 is a central pathological protein in Alzheimer’s disease. The PDB entry 6SZF represents the atomic-resolution structure of amyloid-$\beta$ 42. As is shown in Figure~\ref{fig:protein-structure}a, It is a relatively short single chain protein comprising 185 residues. The protein can undergo self-assembly into $\beta$‑sheet‑rich oligomers and ultimately accumulates as insoluble extracellular amyloid plaques. 
NADPH oxidase 2 is a critical membrane protein complex responsible for reactive oxygen species generation in immune defense. The longer single-chain NOX2 subunit, central to its catalytic activity, was modeled in this study using AlphaFold2, with structural reference derived from the homologous template PDB 3A1F. As illustrated in Figure~\ref{fig:protein-structure}b, the modeled structure comprises approximately 570 residues and adopts a multi-domain architecture featuring six transmembrane helices and cytosolic FAD/NADPH-binding domains.

Hemoglobin is an essential oxygen-transport protein in vertebrates, playing a central physiological role in systemic gas exchange. The PDB entry 1GZX represents the high-resolution structure of human hemoglobin. As shown in Figure~\ref{fig:protein-structure}d, it is a heterotetrameric complex comprising two $\alpha$‑ and two $\beta$‑subunits, each subunit containing a heme prosthetic group that binds oxygen. 
Ecdysone receptor is a key nuclear receptor involved in arthropod molting and metamorphosis. The PDB entry 2HAN represents the DNA-binding domain of the ecdysone receptor. As shown in Figure~\ref{fig:protein-structure}e, It is a multi-chain structure comprising approximately 70 residues. In its DNA-bound state, the domain exhibits a compact $\alpha$-helical fold that specifically recognizes ecdysone response elements. However, when isolated from its cognate DNA (as in the 2HAN structure), the domain displays enhanced conformational flexibility, particularly in its recognition helix and loop regions, making it a suitable model for studying protein dynamics and ligand-induced allostery in a relatively small, well-defined system.
\begin{figure*}[htbp]
\centering
\includegraphics[width=0.8\textwidth]{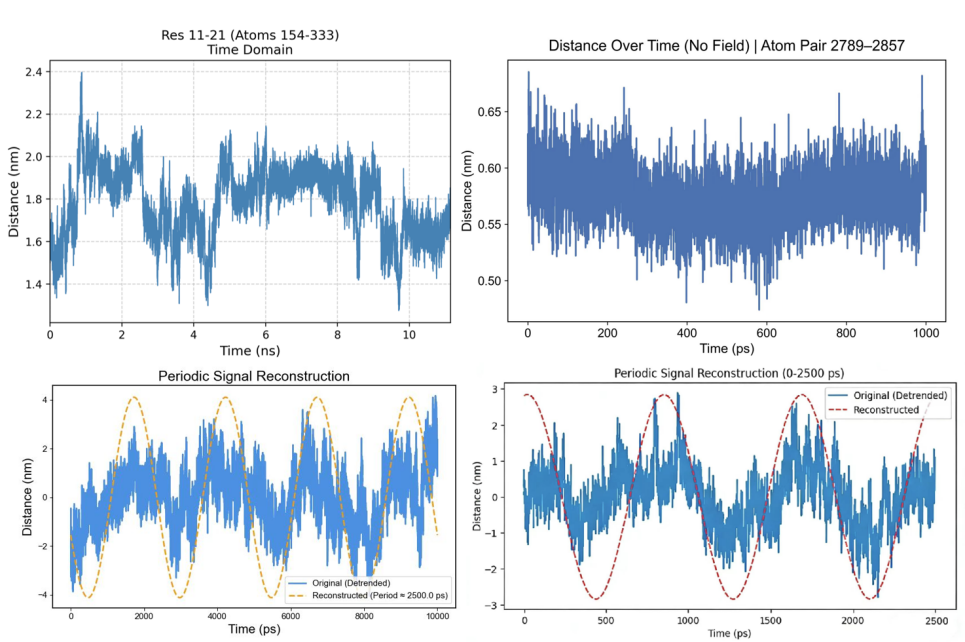}
\caption{Trajectories of selected potential resonant residue or atom pairs. (a) short single-chain protein (PDB: 6SZF), (b) relatively longer single-chain protein (PDB: 3A1F), (c) multichain protein (PDB: 1GZX), (d) Structurally flexible multichain protein (PDB: 2HAN) 
}
\label{fig:pair-distance}
\end{figure*}
In the absence of electromagnetic field, the dynamic trajectories of selected proteins were simulated. Using the screening criteria described in the aforementioned computational methods part, potential resonant residue or atom pairs were selected. As illustrated in the Figure~\ref{fig:pair-distance}, the distance variations between corresponding atoms or residue pairs are depicted, revealing the overall motion and transient characteristics of the proteins. The trajectory encompasses characteristic frequency components dictated by structural rigidity, hydrogen‑bond networks and hydrophobic interactions. 
\begin{figure*}[htbp]
\centering
\includegraphics[width=0.8\textwidth]{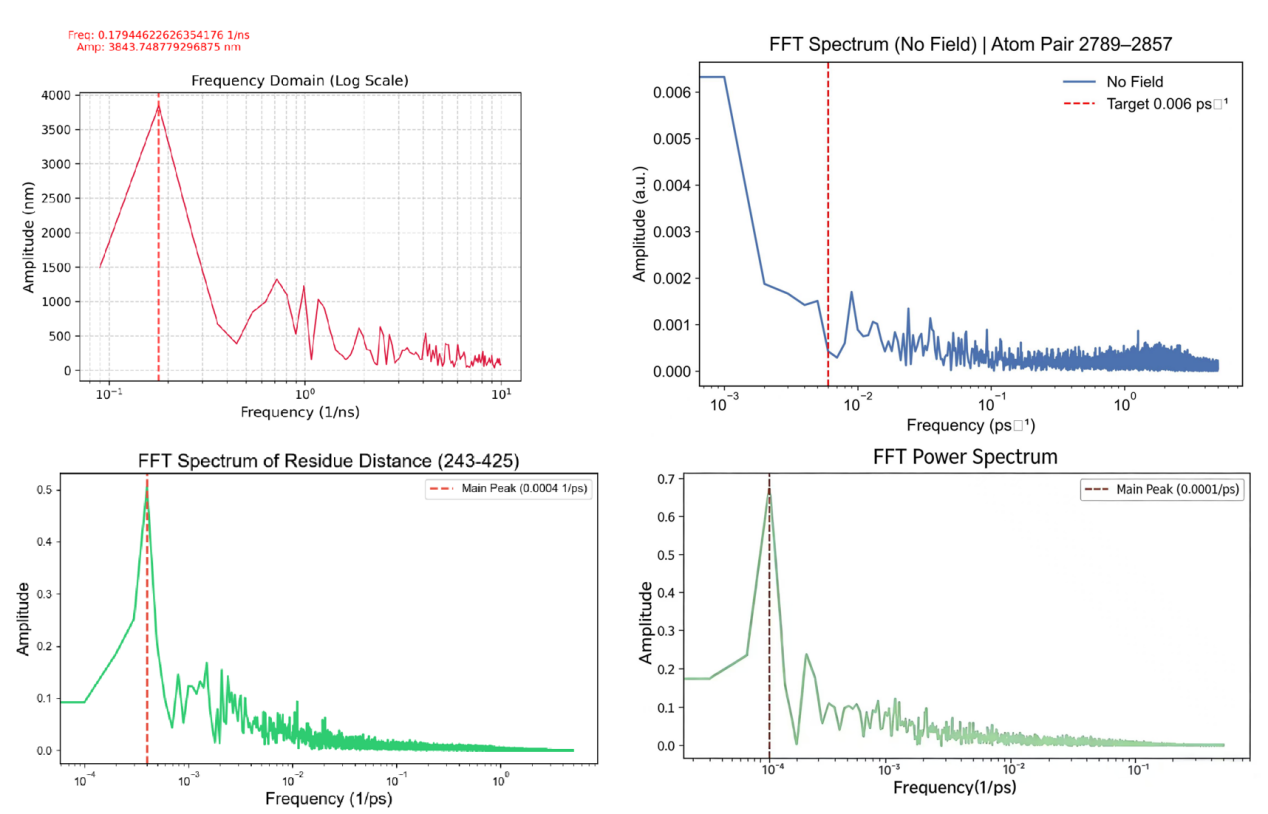}
\caption{Frequency-domain spectral profiles of potential resonant residue/atom pair amplitude dynamics.(a) Residue pairs11-21of protein Amyloid-$\beta$ 42 (PDB: 6SZF), (b) atom pairs of protein NOX2 (PDB: 3A1F), (c) protein (PDB: 1GZX), (d) protein (PDB: 2HAN).
}
\label{fig:fft}
\end{figure*}
To extract these underlying periodic motion modes hidden in the time‑series data in the form of frequency‑amplitude relationships, a Fourier transform was applied to convert the residue‑pair distance data from the time domain to the frequency domain. As shown in Figure~\ref{fig:fft}, dominant frequencies and their intensities were quantitatively extracted, decomposing the complex motion into superpositions of distinct frequency components. The relative weights of different frequency contributions were quantitatively compared via their peak amplitudes.
In the selection of potential resonance frequency, the leftmost and rightmost peaks are excluded due to potential artifacts arising from simulation system effects. Numerical influences such as periodic boundary condition adjustments and long-period oscillations of temperature/pressure coupling in molecular dynamics simulations can generate spurious signals in the extremely low-frequency range. Furthermore, the time scales associated with these frequencies often exceed the total simulation length ( in this study, the total simulation length is 10-20 ns), making it statistically unreliable to sample a full oscillation cycle. Additionally, extremely low frequencies generally correspond to global translational or rotational motions of the protein or slow inter-domain rearrangements, which may be distorted under periodic boundary conditions and exhibit low efficiency in resonant coupling with electromagnetic fields. The rightmost peaks are also disregarded because frequency components near the Nyquist limit are susceptible to aliasing distortion. Extremely high frequencies typically correspond to bond-stretching vibrations (e.g., C–H, O–H) or numerical integration errors. These motions possess high energy barriers and are difficult to couple directly with non-ionizing electromagnetic waves (e.g., microwaves, radio frequency). Moreover, the THz frequency range approaches the infrared spectrum, where the dominant interaction mechanism is thermal rather than resonant coupling.
The characteristic resonance frequencies for the single-chain proteins measured 0.0007 ps-1 and 0.006 ps-1, whereas those for the multichain protein systems are significantly lower at 0.0004 ps-1 and 0.0001 ps-1, respectively. Notably, the resonance frequencies of the single-chain proteins (a and b) are significantly higher than those of their multichain counterparts (c and d). This trend can be understood within the framework of a harmonic approximation~\cite{BrooksKarplus1983,Hinsen2005}, where the frequency:
\begin{equation}
    f \propto \sqrt{\frac{k}{m}}
\end{equation}
For the single‑chain protein (a and b), vibrations are largely confined to local secondary‑structural elements, characterized by a relatively high effective force constant (k) and a small effective mass ( m ), resulting in a higher resonance frequency. In contrast, for the multi-chain proteins (c and d), the dominant vibrational modes correspond to collective, large-amplitude inter‑subunit or interdomain motions. These motions exhibit a significantly larger effective mass and a softer overall force constant arising from ensembles of weak non-covalent interactions, leading to lower resonance frequencies, with the most flexible system (d) showing the lowest value.
\begin{figure*}[htbp]
\centering
\includegraphics[width=0.8\textwidth]{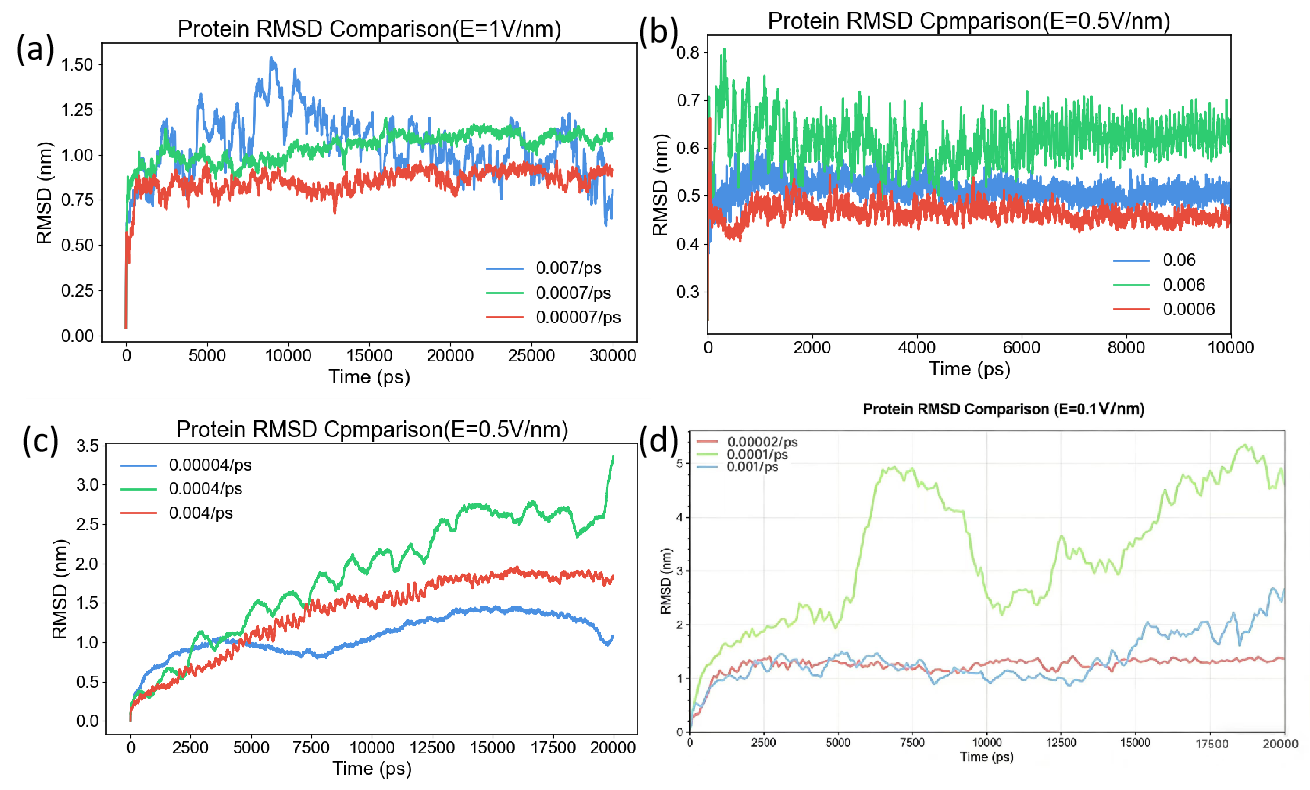}
\caption{RMSD of (a) short single-chain protein (PDB: 6SZF), (b) relatively longer single-chain protein (PDB: 3A1F), (c) multichain protein (PDB: 1GZX), (d) Structurally flexible multichain protein (PDB: 2HAN).
}
\label{fig:rmsd}
\end{figure*}
Under different frequency how much the structure of the protein changed over time is conventionally calculated with root-mean-square deviation (RMSD) against some reference structure (e.g., the equilibrated coordinates or the crystal structure). The RMSD for all residues, backbone, side-chain, and C$\alpha$ atoms can be calculated. Typically, RMSD for all residues, side-chain and C$\alpha$ atoms are often very flexible and whose contributions to RMSD are not necessarily indicative of any real change in the protein structure. Analyzing the backbone RMSD is a direct way of quantifying any deviations in structure of the protein.  
RMSD is the one of the most important parameter to analyze MD trajectory~\cite{Kuzmanic2010}. It can be used to measure the difference between the backbones of a protein from its initial structural conformation to its final position, evaluating the stability of the protein relative to its conformation. RMSD is calculated with respect to the reference native conformation $r_{ref}$  using the following formula:
\begin{equation}
    RMSD(t)=\sqrt{\frac{\sum_{i=1}^{N}m_i\left(r_i(t)-r_i^{ref}\right)^2}{M}}
\end{equation}
where $M=\sum_i m_i$ and $r_i(t)$ represents the atom, i position at the time t after least square fitting the structure to a reference structure.
With an appropriate electric field intensity, an electromagnetic wave at specific frequencies were applied to proteins. The RMSD of the protein backbone under different frequency conditions was compared. The effects of varying electric field strengths on the protein RMSD are provided in the Supporting Information. 
As is shown in Figure~\ref{fig:rmsd}, the conformational perturbation induced by the resonant frequency is markedly stronger than that under off-resonance conditions, and this difference amplifies with increasing structural flexibility. The dynamic RMSD responses of the five selected proteins exhibit a clear gradient. For single‑chain protein (Figure~\ref{fig:rmsd}a and~\ref{fig:rmsd}b), both resonant and off‑resonant RMSD trajectories increase rather gradually, with the smallest difference between the two conditions. For multichain protein (Figure~\ref{fig:rmsd}c and~\ref{fig:rmsd}d), resonant RMSD shows vigorous fluctuations and continuous growth, while off‑resonant RMSD shows minor fluctuations but remains relatively flat, yielding the largest disparity between resonant and non‑resonant responses. 
This gradient in dynamic behavior reflects fundamental differences in the energy‑landscape topography and dissipation pathways of the proteins. Compact single-chain systems possess rugged energy landscapes, where resonant energy is largely dissipated as local fluctuations. In contrast, flexible multichain systems exhibit flatter energy landscapes, allowing resonant energy to drive sustained collective motions that lead to directional conformational drift along specific coordinates. Thus, resonance not only enhances the efficiency of energy coupling but also induces persistent, large-amplitude conformational evolution in flexible multichain architectures. 
Under off-resonance exposure, input energy is rapidly dissipated as incoherent thermal motion. In contrast, resonant excitation enables efficient energy coupling into the collective soft modes of the protein, thereby driving larger‑scale conformational changes. As a result, the backbone root-mean-square deviation (RMSD) is consistently highest at the resonant frequency, an effect that is particularly pronounced in the more loosely organized multichain systems.

\section{Conclusion}
Protein function is governed by both static structure and conformational dynamics, yet traditional experimental techniques cannot fully resolve time-dependent fluctuations. Molecular dynamics simulations address this limitation by tracking atomic motions to identify vibrational modes, dynamic frequencies, and collective motion patterns under physiological conditions, while also enabling investigation of protein responses to external electromagnetic stimuli at high resolution. Current evidence suggests electromagnetic field effects are strongly frequency-dependent, with low-frequency oscillating fields inducing greater conformational alterations than high-frequency fields, though the underlying mechanism remains unclear. Atomistic simulations provide a unique avenue to explore whether specific protein motions resonate with external fields, offering a plausible non-thermal mechanism for field-induced structural modulation. Elucidating these dynamic interaction mechanisms is essential for assessing biological risks of electromagnetic exposure and designing wave-based therapeutic strategies, and integrating molecular dynamics with experimental observations will establish a more comprehensive framework for understanding how external fields modulate protein structure, dynamics, and function.


\bibliography{reference.bib}
\end{document}